\documentclass[onecolumn,showpacs]{revtex4}

\usepackage{graphicx}% Include figure files
\usepackage{dcolumn}% Align table columns on decimal point
\usepackage{bm}% bold math

\input epsf

\begin{document}

\title{\Large Statefinder and {\it Om} Diagnostics for Interacting New Holographic Dark Energy Model
and Generalized Second Law of Thermodynamics}

\author{\bf Ujjal
Debnath$^1$\footnote{ujjaldebnath@yahoo.com ,
ujjal@iucaa.ernet.in} and Surajit
Chattopadhyay$^2$\footnote{surajit$_{_{-}}2008$@yahoo.co.in,
surajit.chattopadhyay@pcmt-india.net}}

\affiliation{$^1$Department of Mathematics, Bengal Engineering and
Science University, Shibpur, Howrah-711 103,
India.\\
$^2$Department of Computer Application (Mathematics Section),
Pailan College of Management and Technology, Bengal Pailan Park,
Kolkata-700 104, India. }

\date{\today}

\begin{abstract}
In this work, we have considered that the flat FRW universe is
filled with the mixture of dark matter and the new holographic
dark energy. If there is an interaction, we have investigated the
natures of deceleration parameter, statefinder and $Om$
diagnostics. We have examined the validity of the first and
generalized second laws of thermodynamics under these interactions
on the event as well as apparent horizon. It has been observed
that the first law is violated on the event horizon. However, the
generalized second law is valid throughout the evolution of the
universe enveloped by the apparent horizon. When the event horizon
is considered as the enveloping horizon, the generalized second
law is found to break down excepting at late stage of the
universe.
\end{abstract}


\maketitle

\section{\normalsize\bf{Introduction}}

In recent observations it is strongly believed that the universe
is experiencing an accelerated expansion. The observation from
type Ia supernovae [1] in associated with Large scale Structure
[2] and Cosmic Microwave Background anisotropies [3] have shown
the evidences to support cosmic acceleration. This observations
lead to a new type of matter which violates the strong energy
condition i.e., $\rho+3p<0$. The energy density of matter
responsible for such a condition to be satisfied at a certain
stage of evolution of the Universe is referred to as dark energy.
This mysterious dark energy with negative pressure leads to this
cosmic acceleration. Also the observations indicate that the
dominating component of the present universe is this dark energy.
Dark energy occupies about 73\% of the energy of our universe,
while dark matter about 23\% and the usual baryonic matter 4\%.
There are several candidates obey the property of dark energy,
given by $-$ quintessence [4], K-essence [5], tachyon [6], phantom
[7], ghost condensate [8,9] and quintom [10], interacting dark
energy models [11], brane world models [12], Chaplygin gas models
[13], agegraphic DE models [14], etc.\\

Recently, a new DE candidate, based on the holographic principle,
was proposed in [15]. By applying the holographic principle to
cosmology, one can obtain the upper bound of the entropy contained
in the universe [16]. Li [17] argued that the total energy in a
region of size $L$ should not exceed the mass of a black hole of
the same size for a system with UV cut-off $\Lambda$. The
holographic dark energy density is defined as [18]
$\rho_{_{\Lambda}}=3c^{2}M_{P}^{2}L^{-2}$, where $M_{P}$ is the
reduced Planck Mass $M_{p}\equiv1/\sqrt{8\pi G}$ and $c$ is a
numerical constant characterizing all of the uncertainties of the
theory, whose value can only be determined by observations.
Recently several works [19] have been done in holographic dark
energy models for acceleration of the universe. Inspired by the
holographic dark energy models, Gao et al [20] proposed to replace
the future event horizon area with the inverse of the Ricci scalar
curvature. So the length $L$ is given by the average radius of
Ricci scalar curvature, $R^{-1/2}$, so that the dark energy
$\rho_{_{\Lambda}}\propto R$, where
$R=-6(2H^{2}+\dot{H}+\frac{k}{a^{2}})$. This model is known as the
\emph{Ricci dark energy model} [20]. They find that this model
works fairly well in fitting the observational data, and it could
also help to understand the coincidence problem. There are few
works on this dark energy model [21]. For purely dimensional
reasons Granda et al [22] proposed a new infrared cut-off for the
holographic density which besides the square of the Hubble scale
also contains time derivative of the Hubble parameter. In favor of
this new term, the underlying origin of the holographic dark
energy is still unknown and that the new term is contained in the
expression for the Ricci scalar which scales as $L^{-1/2}$. So the
holographic density is of the form: $\rho_{_{\Lambda}}\approx
(\alpha H^{2}+\beta \dot{H})$. This is known as the \emph{new
holographic dark energy} [22]. There are some works on the new
version of the holographic dark energy model [23].\\

Bekenstein [24] first considered that there is a relation between
the event of horizon and the thermodynamics of a black hole. In
this case, the event of horizon of the black hole is a measure of
the entropy of it. The thermodynamics in de Sitter space–time was
first investigated by Gibbons and Hawking [25]. In a spatially
flat de Sitter space–time, the event horizon and the apparent
horizon of the Universe coincide and there is only one
cosmological horizon. When the apparent horizon and the event
horizon of the Universe are different, it was found that the first
law and the second law of thermodynamics hold on the apparent
horizon, while they break down if one considers the event horizon
[26]. At the apparent horizon, the first law of thermodynamics is
shown to be equivalent to Friedmann equations [27] if one takes
the Hawking temperature and the entropy on the apparent horizon
and the generalized second law of thermodynamics is obeyed at the
horizon. In this context, there are several studies [28-30] in
thermodynamics for dark energy filled universe on apparent and
event horizons. Here we consider that the flat FRW universe is
filled with the mixture of dark matter and the new holographic
dark energy [22]. If there is an interaction, we shall investigate
the natures of deceleration parameter, statefinder and $Om$
diagnostics during whole evolution of the universe. In the
interaction scenario, the validity of generalized second law
of thermodynamics will be examined on the event horizon. \\

Newness in our study with respect to the earlier works on new
holographic dark energy listed in the references [22] and [23]
lies on the following matters:

\begin{itemize}
    \item Granda and Olivers [23] studied the correspondence of
    new holographic dark energy proposed in [22] with other
    candidates of dark energy like quintessence, tachyon,
    k-essence and dilaton in flat FRW universe. In our study
    instead of studying correspondence between new holographic
    dark energy with other candidates we have considered
    interaction of this dark energy with dark matter and studied
    various diagnostics with zero and non-zero interaction
    parameter $\delta$.
    \item Karami and Fehri [23] generalized the results of Granda
    and Olivers [23] for non-flat universe. They have considered
    non-interacting scenarios. But in our work, we have considered
    interacting scenarios.
    \item Yu et al [23] considered interaction between new
    holographic dark energy and pressureless dark matter. In the
    present paper we have not neglected the pressure of dark
    matter. In our study, we have studied the deceleration, statefinder and
    $Om(z)$ diagnostics under the interactions. Whereas, Yu et al
    [23] studied only the equation of state and fractional energy
    densities. Malekjani et al [23] studied the statefinder
    diagnostics for new holographic dark energy. However, in our
    study we have considered statefinder diagnostics in
    interacting situation.
    \item In none of the studies listed in references [22-23], the laws of
    thermodynamics are investigated. In the present paper we have
    examined the validity of the first and generalized second laws
    of thermodynamics for the universe enveloped by event as well
    as apparent horizon. Moreover, interacting, non-interacting as
    well as single component models have been considered for all
    of the cases.

\end{itemize}

\section{\normalsize\bf{Basic Equations and Solutions}}

The metric of a spatially flat homogeneous and isotropic universe
in FRW model is given by

\begin{equation}
ds^{2}=dt^{2}-a^{2}(t)\left[dr^{2}+r^{2}(d\theta^{2}+sin^{2}\theta
d\phi^{2})\right]
\end{equation}

where $a(t)$ is the scale factor. The Einstein field equations are
given by

\begin{equation}
H^{2}=\frac{1}{3}\rho
\end{equation}
and
\begin{equation}
\dot{H}=-\frac{1}{2}(\rho+p)
\end{equation}

where $\rho$ and $p$ are energy density and isotropic pressure
respectively (choosing $8\pi G=c=1$). Now consider our universe is
filled with dark matter and new holographic dark energy. So we
assume, $\rho=\rho_{_{m}}+\rho_{_{X}}$ and $p=p_{_{m}}+p_{_{X}}$.
Here, $\rho_{_{m}},~p_{_{m}}$ and $\rho_{_{X}},~p_{_{X}}$ are
respectively the energy density and pressure for dark matter and
new holographic dark energy. We know that the dark matter has a
negligible pressure, i.e., $p_{_{m}}\approx 0$. But here we have
taken into account the non-zero value of $p_{_{m}}$ i.e., value of
$p_{m}$ is very small.  \\

The energy conservation equation is given by

\begin{equation}
\dot{\rho}+3\frac{\dot{a}}{a}(\rho+p)=0
\end{equation}

Now we consider the model of interaction between dark matter and
the new holographic dark energy model, through a phenomenological
interaction term $Q$. Keeping into consideration the fact that the
Supernovae and CMB data determine that decay rate should be
proportional to the present value of the Hubble parameter. At this
juncture, it should be sated why we are going for the interaction
between dark energy and dark matter. The models with interaction
between dark energy and dark matter have been studied extensively
in literature [31]. In a study by Pavon and Zimdahl [31] it was
demonstrated that any interaction between pressureless dark matter
with holographic dark energy can solve the coincidence problem.
Some major points favoring the interacting
models are summarized below:\\

  \begin{itemize}
    \item Given the unknown nature of both dark matter and dark energy there is nothing in principle against their mutual interaction (Pavon and Zimdahl [31])
    \item The interacting situation is compatible with SNIa and CMB data (Olivers et al [31], Zhang [31])
    \item Introducing interaction can alleviate coincidence problem (Sheykhi [31], Hu and Ling [31], Sadjadi and Alimohammadi [31])
  \end{itemize}

 In the interacting situation, the two components do not satisfy the
 conservation equation separately. Rather an interaction term $Q$
 is introduced. This interaction term describes the energy flow between
the two fluids. Therefore the conservation equation (4) becomes

\begin{equation}
\dot{\rho}_{_{X}}+3H(\rho_{_{X}}+p_{_{X}})=-Q
\end{equation}
and
\begin{equation}
\dot{\rho}_{_{m}}+3H(\rho_{_{m}}+p_{_{m}})=Q
\end{equation}

There are many different forms of $Q$ in the literature [44].
Following Wei and Cai [44] we choose $Q$ as

\begin{equation}
Q=3\delta H\rho_{_{m}}
\end{equation}

where $\delta$ is the interaction parameter. In principle $\delta$
may be positive or negative. According to Wei and Cai [44] the
cases with positive $\delta$ have physically richer phenomena. If
$\delta=0$ then we get a non-interacting situation. Thus, in the
present study we shall take $\delta\geq0$.
\\

The new holographic dark energy density with infrared cut-off
[22,23] is given by

\begin{equation}
\rho_{_{X}}=3(\alpha H^{2}+\beta \dot{H})
\end{equation}

where $\alpha$ and $\beta$ are constants to be determined. Now
suppose the equation of state for the dark matter is given by
$p_{_{m}}=w_{_{m}}\rho_{_{m}}$, where $w_{_{m}}$ is very small and
so the equation (6) becomes

\begin{equation}
\rho_{_{m}}=\rho_{_{m0}}a^{-3(1+w_{_{m}}-\delta)}
\end{equation}

and consequently equation (2) reduces to

\begin{equation}
H^{2}=C~a^{-\frac{2(\alpha-1)}{\beta}}+\frac{2\rho_{_{m0}}a^{-3(1+w_{_{m}}-\delta)}}{2(1-\alpha)+3\beta(1+w_{_{m}}-\delta)}
\end{equation}

where $\rho_{_{m0}}$ is the present value of matter density (at
$a=1$) and $C$ is the arbitrary integration constant.\\

Now define,

\begin{equation}
\tilde{H}=\frac{H}{H_{0}},
~~~\tilde{\rho}_{_{m}}=\frac{\rho_{_{m}}}{3H_{0}^{2}},~~~\tilde{\rho}_{_{X}}=\frac{\rho_{_{X}}}{3H_{0}^{2}},
~~~\tilde{p}_{_{m}}=\frac{p_{_{m}}}{3H_{0}^{2}},~~~\tilde{p}_{_{X}}=\frac{p_{_{X}}}{3H_{0}^{2}},
~~~\Omega_{m0}=\frac{\rho_{_{m0}}}{3H_{0}^{2}},~~~\Omega_{X0}=\frac{\rho_{_{X0}}}{3H_{0}^{2}},~~~f_{0}=\frac{C}{H_{0}^{2}}
\end{equation}

where $H_{0}$ is the present value of the Hubble parameter,
$\tilde{H}$ is the Hubble expansion rate, current density
parameters are $\Omega_{m0}$ and $\Omega_{X0}$ for matter and dark
energy respectively. So from equation (2), we may conclude that

\begin{equation}
\Omega_{m0}+\Omega_{X0}=1
\end{equation}

Now the relation between scale factor $a$ and the redshift $z$ is
given by

\begin{equation}
a=\frac{1}{1+z}
\end{equation}

Therefore the equations (9), (10), (8) and (5) yield to

\begin{equation}
\tilde{\rho}_{_{m}}=\Omega_{_{m0}}(1+z)^{3(1+w_{_{m}}-\delta)}
\end{equation}

\begin{equation}
\tilde{p}_{_{m}}=w_{_{m}}\Omega_{_{m0}}(1+z)^{3(1+w_{_{m}}-\delta)}
\end{equation}

\begin{equation}
\tilde{H}^{2}=f_{0}(1+z)^{\frac{2(\alpha-1)}{\beta}}+\frac{2\Omega_{_{m0}}(1+z)^{3(1+w_{_{m}}-\delta)}}{2(1-\alpha)+3\beta(1+w_{_{m}}-\delta)}
\end{equation}

\begin{equation}
\tilde{\rho}_{_{X}}=f_{0}(1+z)^{\frac{2(\alpha-1)}{\beta}}+\frac{(2\alpha-3\beta(1+w_{_{m}}-\delta))\Omega_{_{m0}}
(1+z)^{3(1+w_{_{m}}-\delta)}}{2(1-\alpha)+3\beta(1+w_{_{m}}-\delta)}
\end{equation}
and
\begin{equation}
\tilde{p}_{_{X}}=\frac{f_{0}(2(\alpha-1)-3\beta)}{3\beta}(1+z)^{\frac{2(\alpha-1)}{\beta}}+\frac{((2\alpha-3\beta(1+w_{_{m}}-\delta))-2\delta)w_{_{m}}\Omega_{_{m0}}
(1+z)^{3(1+w_{_{m}}-\delta)}}{2(1-\alpha)+3\beta(1+w_{_{m}}-\delta)}
\end{equation}

There are three constants $\alpha$, $\beta$ and $f_{0}$ to be
determined from the above equations. For this purpose let us
suppose that the equation of state for the new holographic dark
energy is $p_{_{X}}=w_{_{X}}\rho_{_{X}}$. Now at present epoch
(i.e., $a=1$, i.e., $z=0$), we have

\begin{equation}
\tilde{\rho}_{_{X0}}=\Omega_{X0},~~~\tilde{p}_{_{X0}}=w_{_{X}}\Omega_{X0}
\end{equation}

So from equations (17), (18) and (19), we obtain

\begin{equation}
f_{0}=1+\frac{2\Omega_{m0}}{3\beta(\delta+w_{_{X}}-w_{_{m}})+\Omega_{m0}(-2+3\beta(w_{_{m}}-w_{_{X}}))}
\end{equation}
and
\begin{equation}
\alpha=1+\frac{3\beta}{2}(1+w_{_{X}})+\left[\frac{3\beta}{2}(w_{_{m}}-w_{_{X}})-1\right]\Omega_{m0}
\end{equation}

So values of $\alpha$ and $f_{0}$ are given in term of the free
parameter $\beta$, but it can be fixed by the behaviour of the
deceleration parameter $q$. The deceleration parameter is given by

\begin{equation}
q=-\frac{a\ddot{a}}{\dot{a}^{2}}=-1-\frac{\dot{H}}{H^{2}}=-1-\frac{a}{2\tilde{H}^{2}}\frac{d\tilde{H}^{2}}{da}
=-1+\frac{(1+z)}{\tilde{H}^{2}}\frac{d\tilde{H}^{2}}{dz}
\end{equation}

Using above results, we obtain

\begin{equation}
q=-\frac{f_{0}(1-\alpha+\beta)(2(1-\alpha)+3\beta(1+w_{_{m}}-\delta))(1+z)^{\frac{2(\alpha-1)}{\beta}}
+\beta(1-3w_{_{m}}+3\delta)\Omega_{m0}(1+z)^{3(1+w_{_{m}}-\delta)}
}{f_{0}\beta(2(1-\alpha)+3\beta(1+w_{_{m}}-\delta))(1+z)^{\frac{2(\alpha-1)}{\beta}}
+2\beta\Omega_{m0}(1+z)^{3(1+w_{_{m}}-\delta)} }
\end{equation}\\

The evolution of deceleration parameter has been shown in figure 1
for $\delta=0$ (red line) and $\delta=.01$ (green line),
~$w_{_{m}}=0.1,~w_{_{X}}=-0.9,~\Omega_{m0}=0.23$ and $\beta=0.5$.
The deceleration parameter decreases from 0.5 to $-1$ for $z$
decreases in the whole evolution. At higher redshifts the
deceleration parameter is positive. This indicates the earlier
decelerating phase of the universe. At lower redshifts, the
deceleration parameter is negative that indicates the accelerated
phase of the universe. The signature flip occurs at $z=0.5$ for
both $\delta=0$ and $\delta=0.01$. Thus, we understand that in the
non-interacting as well as non-interacting situation it is
possible to get the transition from decelerated to accelerated
universe for model under consideration. For the above specified
values,
using equations (20) and (21), we find that $\alpha=1.02$ and $f_{0}=0.71$.\\

\begin{figure}
\includegraphics[height=2.5in]{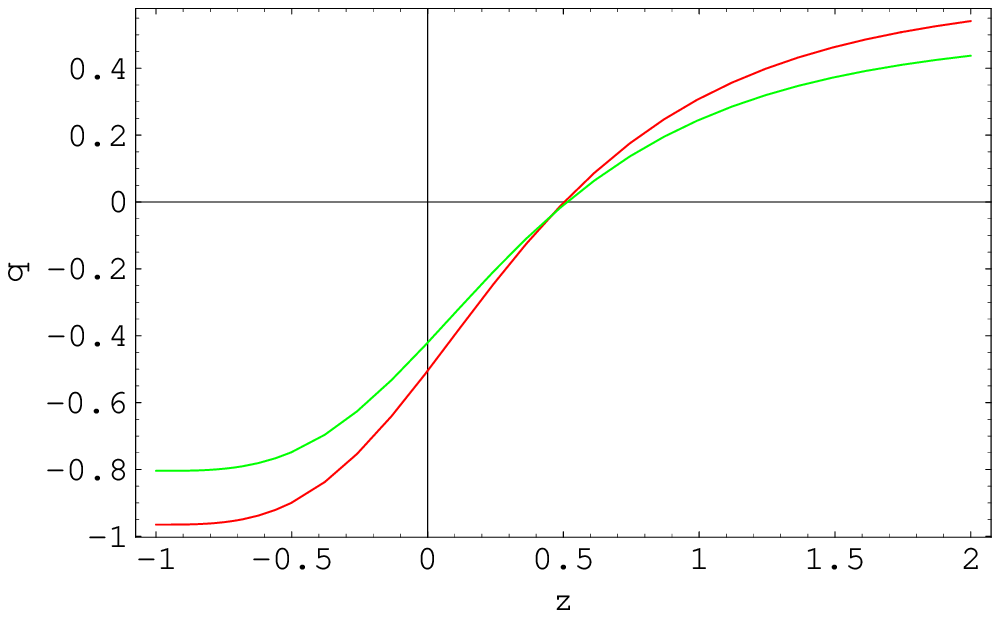}\\
\vspace{1mm} Fig.1\\

\vspace{6mm} Fig. 1 shows the variation of $q$ against redshift
$z$ for $\delta=0$ (red line)$\delta=.01$(green line),
$w_{_{m}}=.1,~w_{_{X}}=-.9,~\Omega_{m0}=.23$ and $\beta=.5$.

 \vspace{6mm}

 \end{figure}

\section{\normalsize\bf{Statefinder Diagnostics}}

Since there are various candidates for the dark energy model, we
often face with the problem of discriminating between them, which
were solved by introducing statefinder parameters. These
statefinder diagnostic pair i.e., $\{r,s\}$ parameters are of the
following form [32]:

\begin{equation}
r=\frac{\dddot{a}}{aH^{3}}
\end{equation}
and
\begin{equation}
s=\frac{r-1}{3\left(q-\frac{1}{2}\right)}
\end{equation}

These parameters are dimensionless and allow us to characterize
the properties of dark energy in a model independent manner. The
statefinder is dimensionless and is constructed from the scale
factor of the Universe and its time derivatives only. The
parameter $r$ forms the next step in the hierarchy of geometrical
cosmological parameters after $H$ and $q$.\\

\begin{figure}
\includegraphics[height=2.5in]{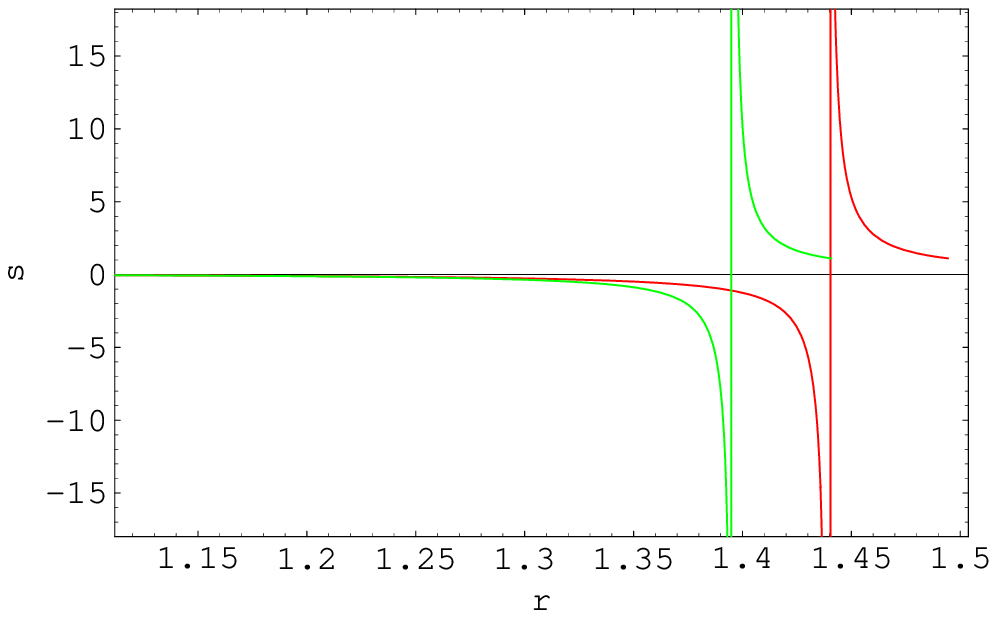}\\
\vspace{1mm} ~~~~~~~~~Fig.2~~~~~~~~~~\\

\vspace{6mm} Fig. 2 shows the variations of $r$ with $s$ for
$\delta=0$ (red line)$\delta=.01$ (green line),
$w_{_{m}}=.1,~w_{_{X}}=-.9,~\Omega_{m0}=.23$ and $\beta=.5$.

\vspace{6mm}

\end{figure}

\begin{figure}
\includegraphics[height=2.5in]{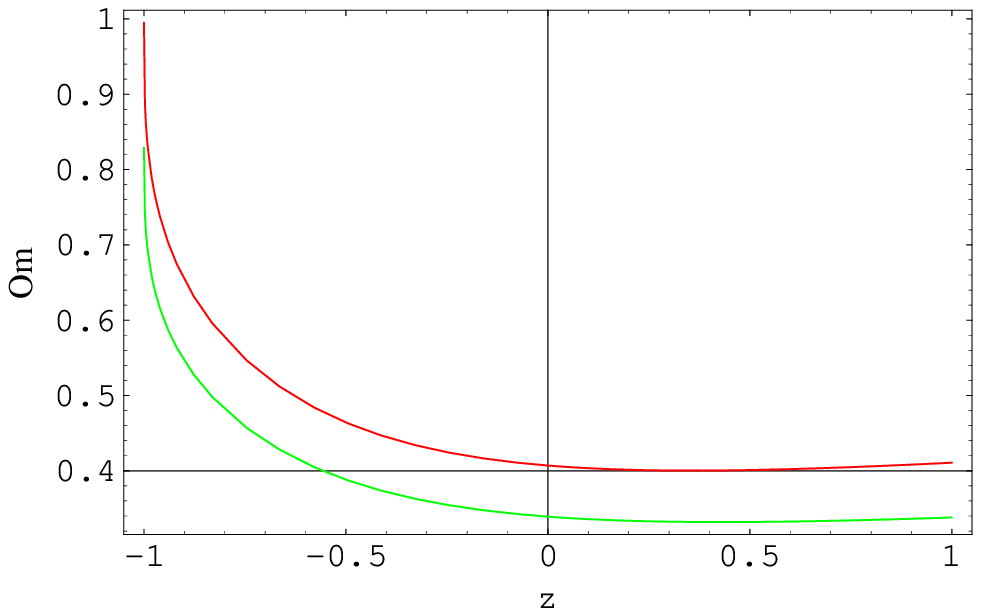}\\
\vspace{1mm} ~~~~~~~~~Fig.3~~~~~~~~~~~\\

\vspace{6mm} Fig. 3 shows the variations of $Om(z)$ against
redshift $z$ for $\delta=0$ (red line)$\delta=.01$, (green
line),~$w_{_{m}}=.1,~w_{_{X}}=-.9,~\Omega_{m0}=.23$ and
$\beta=.5$.

\vspace{6mm}

\end{figure}

Now $r$ and $s$ can be written in terms of Hubble parameter $H$ in
the following forms:

\begin{equation}
r = 1+3\frac{\dot{H}}{H^{2}}+\frac{\ddot{H}}{H^{3}}
\end{equation}
and
\begin{equation}
s = -\frac{3H\dot{H}+\ddot{H}}{3H(2\dot{H}+3H^{2})}
\end{equation}

Or, equivalently [23]

\begin{equation}
r=1+\frac{2a}{\tilde{H}^{2}}\frac{d\tilde{H}^{2}}{da}+\frac{a^{2}}{2\tilde{H}^{2}}\frac{d^{2}\tilde{H}^{2}}{da^{2}}
=1-\frac{(1+z)}{\tilde{H}^{2}}\frac{d\tilde{H}^{2}}{dz}+\frac{(1+z)^{2}}{2\tilde{H}^{2}}\frac{d^{2}\tilde{H}^{2}}{dz^{2}}
\end{equation}
and
\begin{equation}
s=-\frac{4a\frac{d\tilde{H}^{2}}{da}+a^{2}\frac{d^{2}\tilde{H}^{2}}{da^{2}}}{3(3\tilde{H}^{2}+a\frac{d\tilde{H}^{2}}{da})}
=-\frac{2(1+z)\frac{d\tilde{H}^{2}}{dz}-(1+z)^{2}\frac{d^{2}\tilde{H}^{2}}{dz^{2}}}{3(3\tilde{H}^{2}-(1+z)\frac{d\tilde{H}^{2}}{dz})}
\end{equation}

Thus using equation (16), we obtain

\begin{equation}
r=1+\frac{9}{2}(w_{_{m}}-\delta)(1+w_{_{m}}-\delta)+\frac{f_{0}(2(1-\alpha)+3\beta(w_{_{m}}-\delta))
(2(1-\alpha)+3\beta(1+w_{_{m}}-\delta))^{2}(1+z)^{\frac{2(\alpha-1)}{\beta}}
}{2f_{0}\beta^{2}(2(1-\alpha)+3\beta(1+w_{_{m}}-\delta))(1+z)^{\frac{2(\alpha-1)}{\beta}}
+2\beta^{2}\Omega_{m0}(1+z)^{3(1+w_{_{m}}-\delta)} }
\end{equation}
and
\begin{equation}
s=\frac{2f_{0}(\alpha-1)(2-2\alpha+3\beta)(2(1-\alpha)+3\beta(1+w_{_{m}}-\delta))(1+z)^{\frac{2(\alpha-1)}{\beta}}
-18\beta^{2}(w_{_{m}}-\delta)(1+w_{_{m}}-\delta)\Omega_{m0}(1+z)^{3(1+w_{_{m}}-\delta)}
}{3f_{0}\beta(2-2\alpha+3\beta)(2(1-\alpha)+3\beta(1+w_{_{m}}-\delta))(1+z)^{\frac{2(\alpha-1)}{\beta}}
+18\beta^{2}\Omega_{m0}(1+z)^{3(1+w_{_{m}}-\delta)} }
\end{equation}

From (30) and (31), we see that $r$ cannot be expressed explicitly
in terms of $s$. Figure 2 shows the variations of $r$ with $s$ for
$\delta=.01,~w_{_{m}}=.1,~w_{_{X}}=-.9,~\Omega_{m0}=.23$ and
$\beta=.5$. We see that due to evolution of the universe, $r$
decreases and $s$ increases from small positive value to $+\infty$
and the also increases from $-\infty$ to $0$. The section of the
plot with positive $r$ and $s$ gives the radiation phase of the
universe. Then we find that $r$ is finite and
$s\rightarrow-\infty$, which indicates \emph{dust} stage. Also, at
$r=1$, we have $s=0$. This indicates $\Lambda$CDM stage. Thus the
transition from radiation to $\Lambda$CDM stage through dust stage
is obtained under the two-component model with $\delta=0$ and
$\delta=0.01$.
\\

\section{\normalsize $Om$ \bf{Diagnostic}}

As a complementary to $\{r, s\}$, a new diagnostic called $Om$ has
been recently proposed [32], which helps to distinguish the
present matter density contrast $\Omega_{m0}$ in different models
more effectively. The new diagnostic of dark energy $Om$ is
introduced to differentiate $\Lambda$CDM from other DE models. The
starting point for $Om$ diagnostic is the Hubble parameter and it
is defined as [32]

\begin{equation}
Om(x)=\frac{h^{2}(x)-1}{x^{3}-1}
\end{equation}

where $x=z+1$ and $h(x)=\frac{H(x)}{H_{0}}\equiv\tilde{H}$. Thus
$Om$ involves only the first derivative of the scale factor
through the Hubble parameter and is easier to reconstruct from
observational data. For $\Lambda CDM$ model, $Om = \Omega_{0m}$ is
a constant, independent of redshift $z$. It provides a null test
of cosmological constant. The benefit for $Om$ diagnostic is that
the quantity $Om$ can distinguish DE models with less dependence
on matter density $\Omega_{0m}$ relative to the EOS of DE
$w_{_{X}}$ [32]. $Om$ and statefinder diagnostics for GCG model
and decaying vacuum model from cosmic observations has also been
discussed in [33].\\

Now in our interacting new holographic dark energy model, we
obtain

\begin{equation}
Om(z)=\frac{\tilde{H}^{2}(z)-1}{(1+z)^{3}-1}=\frac{f_{0}(1+z)^{\frac{2(\alpha-1)}{\beta}}+
\frac{2\Omega_{m0}(1+z)^{3(1+w_{_{m}}-\delta)}}{2(1-\alpha)+3\beta(1+w_{_{m}}-\delta)}
-1 }{(1+z)^{3}-1}
\end{equation}

In figure 3, we plot the evolution of $Om(z)$ against redshift $z$
corresponding to $\delta=0$ (red line), $\delta=.01$, (green
line),~$w_{_{m}}=.1,~w_{_{X}}=-.9,~\Omega_{m0}=.23$ and
$\beta=.5$. It may be seen that $Om(z)$ increases as $z$
decreases, so $Om(z)$ increases due to evolution of the universe
in the interacting model. According to reference [32], positive
slope of $Om(z)$ suggests phantom $(w<-1)$ and negative slope of
$Om(z)$ suggests quintessence $(w>-1)$. In figure 3 we find that
the $Om(z)$ is characterized negative slope indicating
quintessence like behavior in the presence of interaction as well
as non-interaction.
\\

\section{\normalsize\bf{First law of thermodynamics}}

In this section we are going to examine the validity of the first
law of thermodynamics on the event as well as on the apparent
horizon in the interacting situation under consideration. The
first law of thermodynamics to the apparent horizon in the FRW
universe has been studied in the reference [34]. The
thermodynamics of the de Sitter universe was considered in the
reference [35], where it was shown that de Sitter universe
experiences accelerated expansion and has only one cosmological
horizon analogous to the black hole horizon. In reference [36] it
was disclosed that the first law of thermodynamics holds in the
physically relevant part of the accelerating universe enveloped by
the dynamical apparent horizon, while does not hold in the region
enveloped by the cosmological event horizon. First we examine the
validity of the first law of thermodynamics on the event horizon
whose radius is given by [26]

\begin{equation}
R_{h}=a\int_{t}^{\infty}\frac{dt}{a}=-\frac{1}{(1+z)H_{0}}\int_{z}^{-1}\frac{dz}{\tilde{H}}
\end{equation}

Differentiating with respect to cosmic time $t$, we obtain

\begin{equation}
\dot{R}_{h}=HR_{h}-1
\end{equation}

The temperature and the entropy on the event horizon are given as
[28,34]

\begin{equation}
T_{h}=\frac{1}{2\pi R_{h}},~~~~~~~S_{h}=\frac{\pi
R_{h}^{2}}{G}=8\pi^{2}R_{h}^{2}~,~~~~~ (8\pi G=1)
\end{equation}

The amount of the energy crossing on the event horizon is

\begin{equation}
-dE_{h}=4\pi R_{h}^{3}HT_{\mu\nu}k^{\mu}k^{\nu}dt=-8\pi R_{h}^{3}
H\dot{H}dt
\end{equation}

Validity of the first law of thermodynamics means validity of the
equation [34]

\begin{equation}
T_{h}dS_{h}=-dE_{h}=-8\pi R_{h}^3 H\dot{H}dt
\end{equation}

In the present interaction, the Hubble's parameter $H$ is of the
form obtained in equation (10). In figure 4 we present
$T_{h}\dot{S}_{h}+ 8\pi R_{h}^{3} H\dot{H}$ against redshift $z$
when the universe if filled with only new holographic dark energy.
It is observed in this figure that the said quantity is staying at
negative level throughout the evolution of the universe. This
indicates that the first law of thermodynamics is not valid on the
event horizon when the universe is considered to be filled with
only new holographic dark energy. In figure 5 we plot
$T_{h}\dot{S}_{h}+ 8\pi R_{h}^{3} H\dot{H}$ against redshift $z$
when there is an interaction between new holographic dark energy
and dark matter. Here also we observe that the first law of
thermodynamics fails to be valid on the event horizon.\\

Now we examine the validity of the first law of thermodynamics on
the apparent horizon. Radius of the apparent horizon for the FRW
universe is given by

\begin{equation}
R_{a}=\frac{1}{\sqrt{H^{2}+\frac{k}{a^{2}}}}
\end{equation}

As we are considering flat FRW universe, $k=0$ and consequently
$R_{a}=1/H$. On the apparent horizon, temperature $T_{a}=H/2\pi$
and entropy $S_{a}=8 \pi^{2}/H^{2}$. The obvious consequence is

\begin{equation}
T_{a}\dot{S}_{a}+ 8\pi R_{a}^3 H\dot{H}=0
\end{equation}

Thus, the first law of thermodynamics is always valid on the
apparent horizon irrespective of the interaction. Thus, the above
revelations are consistent with the observations made in the
reference [36].\\

\begin{figure}
\includegraphics[height=2in]{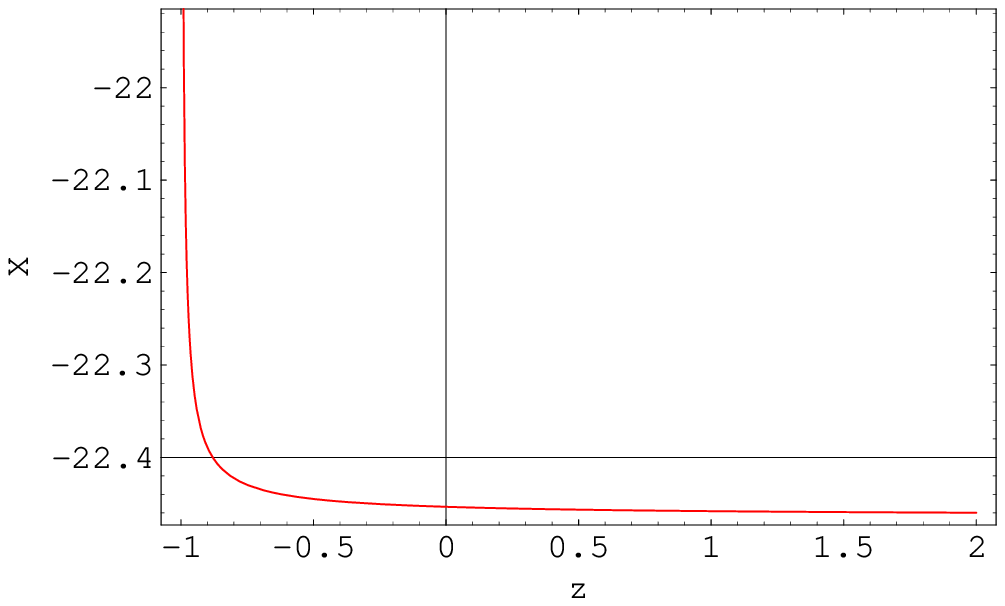}~~~~
\includegraphics[height=2in]{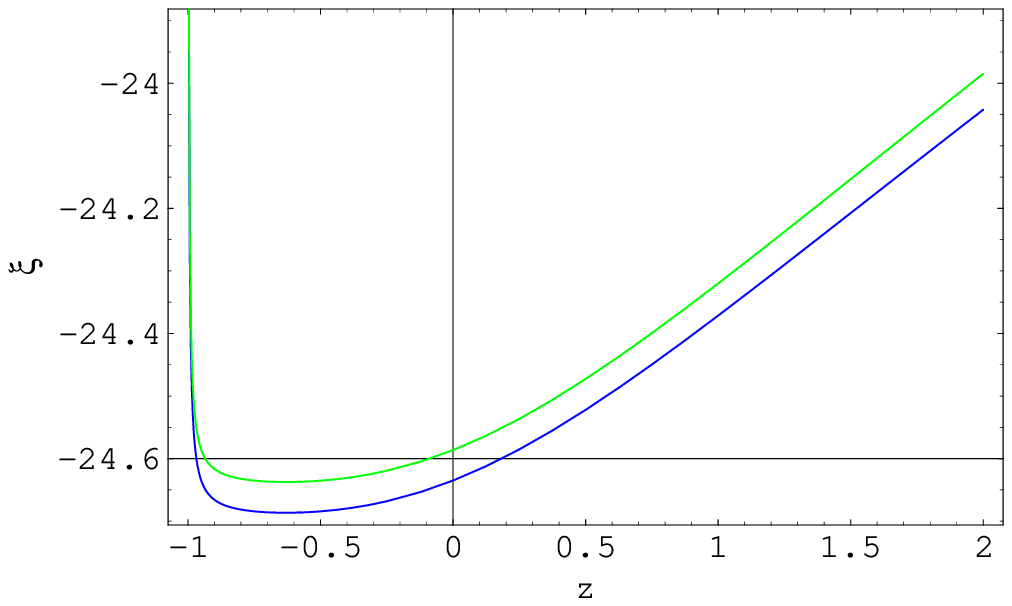}\\
\vspace{1mm} ~~~~~~~Fig.4~~~~~~~~~~~~~~~~~~~~~~~~~~~~~~~~~~~~~~~~~~~~~~~~~~~~~~~~~~~~~~~~~~~~~~~~~Fig.5~~~\\

\vspace{6mm} Fig. 4 shows the variations of $X=T_{h}\dot{S}_{h}+
8\pi R_{h}^{3} H\dot{H}$ against redshift $z$ for only new
holographic dark energy model (without dark matter) and fig. 5
shows the variations of $X=T_{h}\dot{S}_{h}+ 8\pi R_{h}^3
H\dot{H}$ against redshift $z$ for $\delta=0$ (blue line),
$\delta=.01$ (green
line),~$w_{_{m}}=.1,~w_{_{X}}=-.9,~\Omega_{m0}=.23$ and
$\beta=.5$.\\

 \vspace{6mm}

 \end{figure}

\section{\normalsize\bf{Generalized second law of thermodynamics}}

Importance of examining the validity of the generalized second law
of thermodynamics in the accelerating universe driven by dark
energy have been emphasized by a plethora of literatures [37-42].
Benkenstein [24] assumed that there is a relation between the
event horizon and the thermodynamics of a black hole and
consequently the second law of thermodynamics was modified to the
generalized second law of thermodynamics. In references like [38]
and [39], the validity of the generalized second law of
thermodynamics has been studied on the event horizon. According to
the generalized second law, entropy of matter and fluids inside
the horizon plus the entropy of the horizon do not decrease with
time [37, 38]. Using a specific model of dark energy, the
generalized second law as defined in the region enveloped by the
apparent horizon as well as in the event horizon was examined in
[43], where it was found that it is obeyed in the case of the
universe enveloped by the apparent horizon, not on the event
horizon. In this section, we shall examine the validity of the
generalized second law for both event and apparent horizon. \\

 First we consider the
validity of the generalized second law of thermodynamics
considering the event horizon as the enveloping horizon of the
universe. To show the validity of the generalized second law of
thermodynamics we start with Gibb's equation [26, 39]

\begin{equation}
T_{h}dS=pdV+d(\rho V)
\end{equation}

where, the volume of the sphere is $V=\frac{4}{3}\pi R_{h}^{3}$.
Using (34) - (37), the rate of change of total entropy is obtained
as

\begin{equation}
\dot{S}_{total}=\dot{S}+\dot{S}_{h}
=16\pi^{2}R_{h}\left[-(1+z)H_{0}^{2}R_{h}^{2}\frac{d\tilde{H}}{dz}+H_{0}\tilde{H}R_{h}-1\right]
\end{equation}

The generalized second law will be valid if

\begin{equation}
\dot{S}+\dot{S}_{h}\ge 0
~~~i.e.,~~(1+z)H_{0}^{2}R_{h}^{2}\frac{d\tilde{H}}{dz} \ge
H_{0}\tilde{H}R_{h}-1
\end{equation}

\begin{figure}
\includegraphics[height=2in]{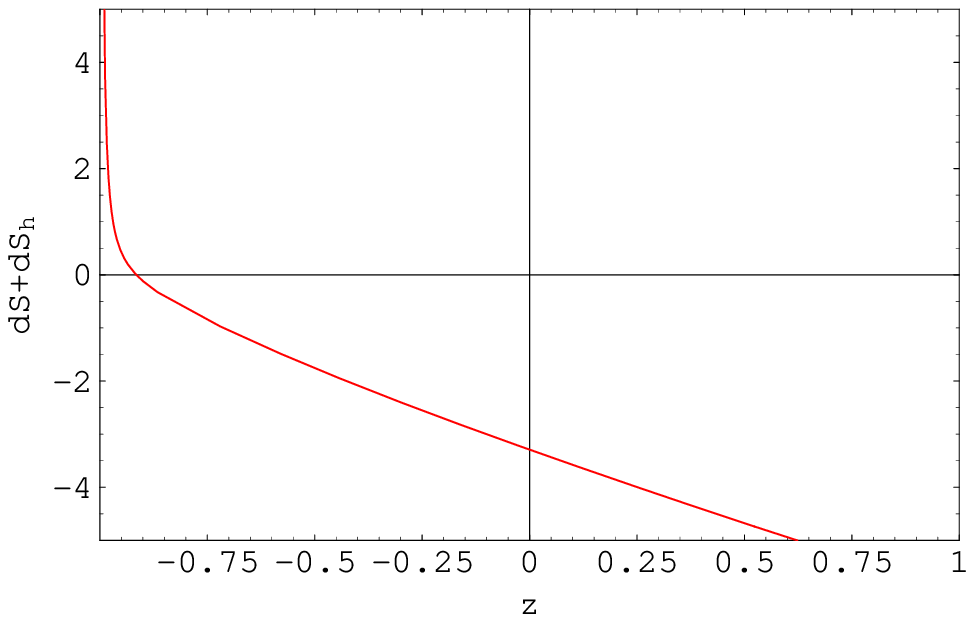}~~~~
\includegraphics[height=2in]{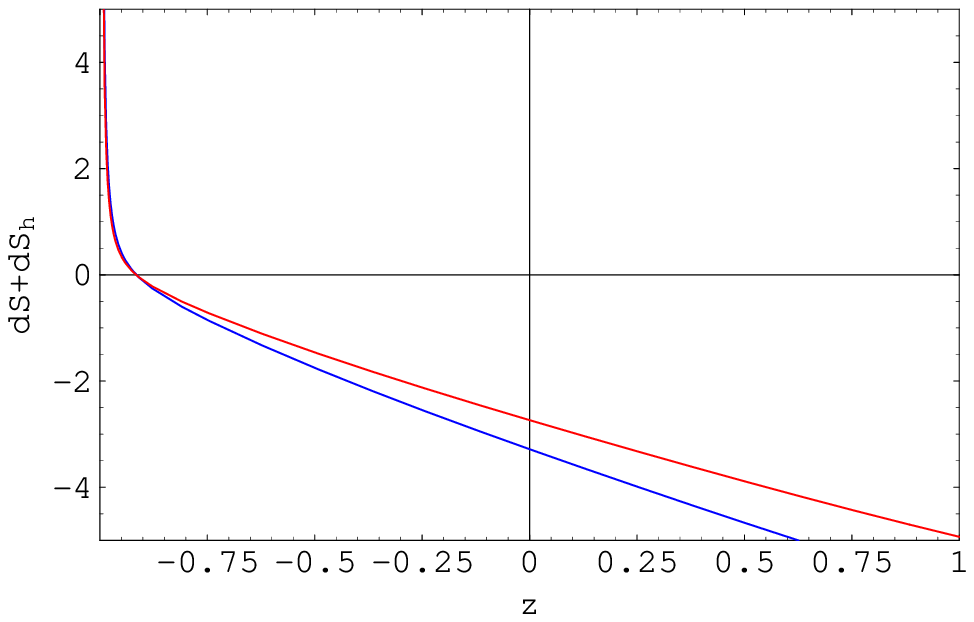}\\
\vspace{1mm} ~~~~~~~Fig.6~~~~~~~~~~~~~~~~~~~~~~~~~~~~~~~~~~~~~~~~~~~~~~~~~~~~~~~~~~~~~~~~~~~~~~~~~Fig.7~~~\\

\vspace{6mm} Fig. 6 shows the variations of the time derivative of
total entropy against redshift $z$ for only new holographic dark
energy model (without dark matter) and fig. 7 shows the variations
of the time derivative of total entropy against redshift $z$ for
$\delta=0$ (blue line), $\delta=.01$ (red
line),~$w_{_{m}}=.1,~w_{_{X}}=-.9,~\Omega_{m0}=.23$ and
$\beta=.5$. In these figures the entropies are calculated for the
universe enveloped by the event horizon.\\

 \vspace{6mm}

 \end{figure}

\begin{figure}
\includegraphics[height=2in]{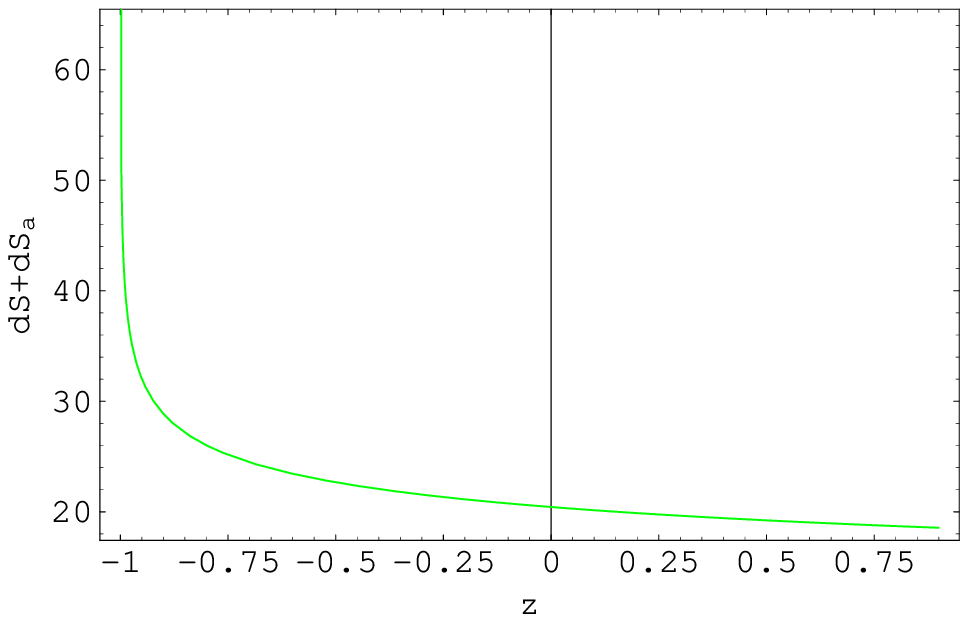}~~~~
\includegraphics[height=2in]{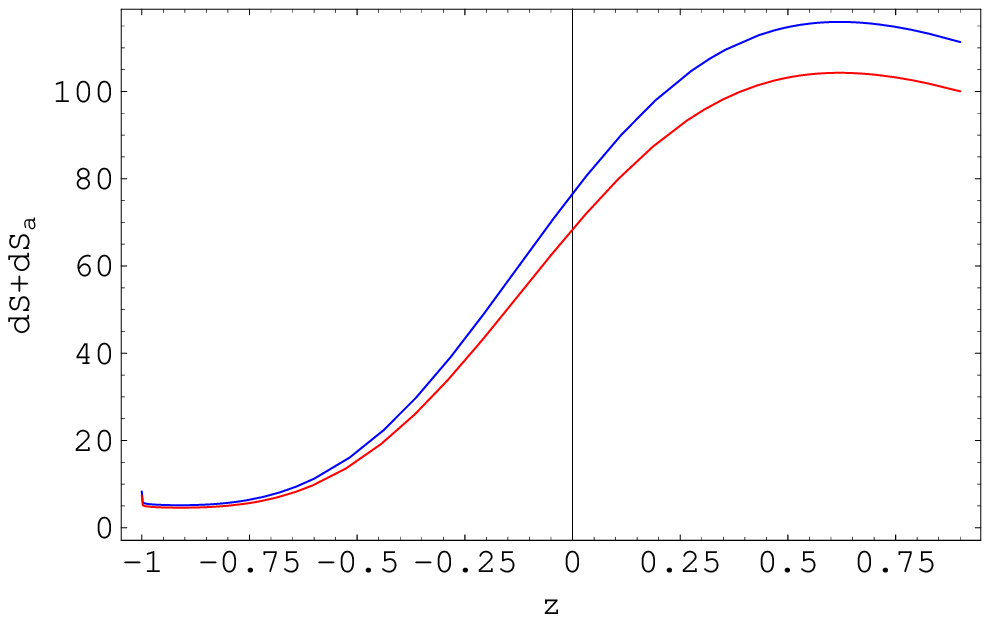}\\
\vspace{1mm} ~~~~~~~Fig.8~~~~~~~~~~~~~~~~~~~~~~~~~~~~~~~~~~~~~~~~~~~~~~~~~~~~~~~~~~~~~~~~~~~~~~~~~Fig.9~~~\\

\vspace{6mm} Fig. 8 shows the variations of the time derivative of
total entropy against redshift $z$ for only new holographic dark
energy model (without dark matter) and fig. 9 shows the variations
of the time derivative of total entropy against redshift $z$ for
$\delta=0$ (blue line), $\delta=.01$ (red
line),~$w_{_{m}}=.1,~w_{_{X}}=-.9,~\Omega_{m0}=.23$ and
$\beta=.5$. In these figures the entropies are calculated for the
universe enveloped by the apparent horizon.\\

 \vspace{6mm}

 \end{figure}

In figure 6 we have plotted $\dot{S}+\dot{S}_{h}$ against redshift
$z$ considering that the universe is filled with the new
holographic dark energy only i.e. there is no dark matter. In this
situation we find that the $\dot{S}_{total}$ is positive at late
stage of the universe i.e. at low redshift. In figure 7 we have
plotted $\dot{S}_{total}$ against $z$ in the interacting
$(\delta\neq0)$ as well as non-interacting $(\delta=0)$
situations. In both of the cases it is observed that the
generalized second law of thermodynamics breaks down excepting
very late stage of the universe.\\

Next we consider the generalized second law of thermodynamics in
the universe enveloped by apparent horizon $R_{a}=1/H$. Time
derivatives on and within the apparent horizon are given by

\begin{equation}
\dot{S}_{a}=-\frac{16\pi\dot{H}}{H^{3}}~;\\~~~~~~~\dot{S}=\frac{3\pi}{2H}(1+3w_{x}\Omega_{x})(1+w_{x}\Omega_{x})
\end{equation}

So the rate of change of total entropy of the universe bounded by
apparent horizon is given by

\begin{equation}
\dot{S}_{total}=\dot{S}+\dot{S}_{a}=\frac{3\pi}{2H}(1+3w_{x}\Omega_{x})(1+w_{x}\Omega_{x})-\frac{16\pi\dot{H}}{H^{3}}
\end{equation}

In figure 8 we have plotted $\dot{S}+\dot{S}_{A}$ against redshift
$z$ considering that the universe is filled with the new
holographic dark energy only i.e. there is no dark matter. In
figure 9 we have plotted $\dot{S}_{total}$ against $z$ in the
interacting $(\delta\neq0)$ as well as non-interacting
$(\delta=0)$ situations. In these situations we find that the
$\dot{S}_{total}$ is always positive in the evolution of the
universe. In both of the cases it is observed that the generalized
second law of thermodynamics is always satisfied.\\

\section{\normalsize\bf{Discussions}}

In this work, we have considered that the flat FRW universe is
filled with the mixture of dark matter and the new holographic
dark energy proposed by [22]. We have considered here that there
is an interaction between dark matter and dark energy. The
solutions have been obtained in terms of redshift for a particular
form of interaction term. The expressions of $\alpha$ and $f_{0}$
are obtained in term of the free parameter $\beta$, but it can be
fixed by the behaviour of the deceleration parameter $q$. Also in
the current observations, for specific values of other parameters,
we find that $\alpha=1.02$ and $f_{0}=0.71$. If there is an
interaction, we have investigated the natures of deceleration
parameter, statefinder and $Om$ diagnostics during whole evolution
of the universe (figures 1-3). At the same time we have examined
the cases of the interaction parameter $\delta=0$ which implies
the existence dark matter and dark energy in the form of a mixture
without interaction. It has been observed that the nature of the
various diagnostic parameters are not significantly influenced by
the interaction. It has been observed from the $r$-$s$ plot that
the two-component model is able to explain the evolution of the
universe from radiation to $\Lambda$CDM though dust stage. The
negative slope of the $Om(z)$ has indicated that the model behaves
like quintessence. A signature flip for the deceleration parameter
$q$ has occurred thereby indicating the evolution of the universe
from early deceleration to present acceleration.\\

In the next phase we have investigated the validity of the laws of
thermodynamics considering apparent as well as event horizon as
the enveloping horizon of the universe. When the new holographic
dark energy is considered alone (without dark matter), it is
observed that first law of thermodynamics is violated on the event
horizon (figure 4) throughout the evolution of the universe.
However, for this case the generalized second law of
thermodynamics is valid in the late stage of the universe
enveloped by the event horizon (figure 6). When the apparent
horizon is considered as the enveloping horizon of the universe,
the new holographic dark energy (without dark matter) satisfies
the generalized second law throughout the evolution of the
universe (figure 8). While considering the new holographic dark
energy with dark matter we find that for $\delta\neq0$, the first
law breaks down throughout the evolution of the universe enveloped
by the event horizon (figure 5) and generalized second law ia
valid only in the late stage of the universe enveloped by the
event horizon (figure 7). However, figure 9 shows that the
generalized second law is valid when we assume that the universe
is enveloped by the apparent horizon. Similar behaviors are
discerned for non-interacting $(\delta=0)$ situations also. We
have noted from figures 8 and 9 that the time derivative of total
entropy has an increasing pattern for only dark energy model with
the evolution of the universe enveloped by the apparent horizon,
whereas for the two-component model is is decaying from higher to
lower redshifts. However, when the event horizon is assumed as the
enveloping horizon, the two and single-component models behave
similarly for the time derivative of total entropy. \\\\

{\bf Acknowledgement:}\\

One of the author (UD) is thankful to IUCAA, Pune, India for
providing Associateship Programme under which the part of the work
was carried out.\\

{\bf References:}\\\\\
[1] A. G. Riess etal, \textit{Astron. J.} \textbf{116} 1009
(1988): S. Perlmutter etal, \textit{Astrophys. J.} \textbf{517}
565 (1999).\\\
[2] K. Abazajian etal, \textit{Astron. J.} \textbf{128} 502
(2004); Abazajian etal, \textit{Astron. J.} \textbf{129} 1755
(2005).\\\
[3] D. N. Spergel etal, \textit{Astrophys. J. Suppl.} \textbf{148}
175 (2003);
\textit{Astrophys. J. Suppl.} \textbf{170} 377 (2007). \\\
[4] P. J. E. Peebles and B. Ratra, \textit{Astrophys. J.}
\textbf{325} L17 (1988); R. R. Caldwell, R. Dave and P. J.
Steinhardt, \textit{Phys. Rev. Lett.} \textbf{80} 1582
(1998).\\\
[5] C. Armendariz - Picon, V. F. Mukhanov and P. J. Steinhardt,
\textit{Phys. Rev. Lett.} \textbf{85} 4438
(2000).\\\
[6] A. Sen, \textit{JHEP} \textbf{0207} 065 (2002).\\\
[7] R. R. Caldwell, \textit{Phys. Lett. B} \textbf{545} 23
(2002).\\\
[8] N. Arkani-Hamed, H. C. Cheng, M. A. Luty and S. Mukohyama,
\textit{JHEP} \textbf{0405} 074
(2004).\\\
[9] F. Piazza and S. Tsujikawa, \textit{JCAP} \textbf{0407} 004
(2004).\\\
[10] B. Feng, X. L. Wang and X. M. Zhang, \textit{Phys. Lett. B}
\textbf{607} 35 (2005); Z. K. Guo, Y. S. Piao, X. M. Zhang and Y.
Z. Zhang, \textit{Phys. Lett. B} \textbf{608} 177 (2005).\\\
[11] L. Amendola, \textit{Phys. Rev. D} \textbf{62} 043511 (2000);
X. Zhang, \textit{Mod. Phys. Lett. A} \textbf{20} 2575
(2005).\\\
[12] V. Sahni and Y. Shtanov, \textit{JCAP} \textbf{0311} 014
(2003).\\\
[13] A. Y. Kamenshchik, U. Moschella and V. Pasquier,
\textit{Phys. Lett. B} \textbf{511} 265
(2001).\\\
$[14]$ R. G. Cai, {\it Phys. Lett. B} {\bf 657} 228 (2007); H. Wei
and R. G. Cai, {\it Phys. Lett. B} {\bf 660} 113 (2008).\\
$[15]$ P. Horava and D. Minic, {\it Phys. Rev. Lett.} {\bf 85}
1610 (2000); P. Horava and D. Minic, {\it Phys. Rev. Lett.} {\bf
509} 138 (2001); S. Thomas, {\it Phys. Rev. Lett.} {\bf 89} 081301
(2002).\\
$[16]$ W. Fischler and L. Susskind, {\it hep-th}/9806039.\\
$[17]$ M. Li, {\it Phys. Lett. B} {\bf 603} 1 (2004).\\
$[18]$ X. Zhang, {\it Int. J. Mod. Phys. D} {\bf 14} 1597 (2005).\\
$[19]$ W. Fischler and L. Susskind, {\it hep-th}/9806039; Y. Gong,
{\it Phys. Rev. D} {\bf 70} 064029 (2004); B. Wang, Y. Gong and E.
Abdalla, {\it Phys. Lett. B} {\bf 624} 141 (2005); X. Zhang, {\it
Int. J. Mod. Phys. D} {\bf 14} 1597 (2005); D. Pavon and W.
Zimdahl, {\it hep-th}/0511053; A. G. Cohen et al., {\it Phys. Rev.
Lett.} {\bf 82} 4971 (1999).\\
$[20]$ C. Gao, F. Wu and X. Chen, {\it Phys. Rev. D} {\bf 79}
043511 (2009).\\
$[21]$ C. -J. Feng, {\it Phys. Lett. B} {\bf 670} 231 (2009); L.
Xu, J. Lu and W. Li, ; {\it Eur. Phys. J. C} {\bf 64} 89 (2009);
L. Xu, W. Li and J. Lu, {\it Mod. Phys. Lett. A} {\bf 24} 1355
(2009); C. -J. Feng and X. -Z. Li, {\it Phys. Lett. B} {\bf 680}
355 (2009); C. -J. Feng, {\it Phys. Lett. B} {\bf 672} 94 (2009);
C. -J. Feng, hep-th/0806.0673; K. Y. Kim, H. W. Lee and Y. S.
Myung, arXiv:0812.4098v1 [gr-qc]; M. Suwa and T. Nihei, arXiv:0911.4810v1 [astro-ph].\\
$[22]$ L. N. Granda and A. Oliveros, {\it Phys. Lett. B} {\bf 669} 275 (2008).\\
$[23]$ M. Malekjani, A. Khodam-Mohammadi, N. Nazari-pooya, {\it
Astrophys. Space Sci.} {\bf 332} 515 (2010);  L. N. Granda and A.
Oliveros, {\it Phys. Lett. B} {\bf 671} 199 (2009); K. Karami and
J. Fehri,
{\it Int. J. Theor. Phys.} {\bf 49} 1118 (2010); F. Yu, J. Zhang, J. Lu, W. Wang and Y. Gui, {\it Phys. Lett. B} {\bf 688} 263 (2010).\\
$[24]$ J. D. Bekenstein, {\it Phys. Rev. D} {\bf 7} 2333 (1973).\\
$[25]$ G. W. Gibbons and S. W. Hawking, {\it Phys. Rev. D} {\bf 15} 2738 (1977).\\
$[26]$ B. Wang, Y. G. Gong and E. Abdalla, {\it Phys. Rev. D} {\bf 74} 083520 (2006).\\
$[27]$ R. G. Cai and S. P. Kim, {\it JHEP} {\bf 02} 050 (2005).\\
$[28]$ Y. Gong, B. Wang and A. Wang, {\it JCAP} {\bf 01} 024
(2007); T. Padmanabhan, {\it Class. Quantum Grav.} {\bf 19} 5387
(2002); R. -G. Cai and N. Ohta, {\it Phys. Rev. D} {\bf 81}
084061; R. G. Cai and L. -M. Cao, {\it Nucl. Phys. B} {\bf 785}
135 (2007); M. Akbar and R. -G. Cai, {\it Phys. Lett. B} {\bf 635}
7 (2006); R. -G. Cai, L. -M. Cao, Y. -P. Hu and S. P. Kim, {\it
Phys. Rev. D} {\bf 78} 124012 (2008).\\
$[29]$ H. M. Sadjadi, {\it Phys. Lett. B} {\bf 645} 108 (2007); H.
M. Sadjadi, {\it Phys. Rev. D} {\bf 76} 104024 (2007); M. Jamil,
A. Sheykhi and M. U. Farooq, arXiv:1003.2093[hep-th]; S. -F. Wu,
B. Wang and G. -H Yang, {\it Nucl. Phys. B} {\bf 799} 330
(2008).\\
$[30]$ M. R. Setare and S. Shafei, {\it JCAP} {\bf 09} 011 (2006);
M. R. Setare, {\it Phys. Lett. B} {\bf 641} 130 (2006); P. C. W.
Davies, {\it Class. Quant. Grav.} {\bf 4} L225 (1987); N. Mazumder
and S. Chakraborty, arXiv: 1003.1606[gr-qc]; U. Debnath, arXiv: 1006.2217[gr-qc].\\
$[31]$ X. Zhang, arXiv:hep-ph/0410292v1; B. Hu and Y. Ling, {\it
Phys. Rev. D} {\bf 73} 123510 (2006); H. M. Sadjadi and M.
Alimohammadi, {\it Phys. Rev. D} {\it 74} 103007 (2006); H. Wei
and R-G. Cai, {\it Euro. Phys. Journal C} {\bf 59} 99 (2009); A.
Sheykhi, {\it Phys. Rev. D} {\bf 81} 023525 (2010); H. Wei and
R-G. Cai, {\it Phys. Lett. B} {\bf 655} 1 (2007); D. Pavon and W.
Zimdahl, {\it Phys. Lett. B} {\bf 628} 206 (2005); G. Olivers, F.
Atrio and D. Pavon, {\it Phys. Rev. D} {\bf 71} 063523 (2005).
\\
$[32]$ V. Sahni, T. D. Saini, A. A. Starobinsky and U. Alam, {\it
JETP Lett.} {\bf 77} 201 (2003); V. Sahni, A. Shafieloo and  A. A. Starobinsky, {\it Phys. Rev. D} {\bf 78}, 103502 (2008).\\
$[33]$ J. Lu, L. Xu, Y. Gui and B. Chang, arXiv:0812.2074v2
[astro-ph]; M. L. Tong and Y. Zhang, {\it Phys. Rev. D} {\bf 80} 023503 (2009).\\
$[34]$ R-G. Cai and P. K. Song, {\it JHEP} {\bf 02} 50 (2005); R. Bousso, {\it Phys. Rev. D} {\bf 71} 064024 (2005).\\
$[35]$ G. W. Gibbons and S. W. Hawking, {\it Phys. Rev. D} {\bf 15} 2738 (1977).\\
$[36]$ B. Wang, Y. Gong and E. Abdalla, {\it Phys.Rev. D} {\bf 74} 083520 (2006).\\
$[37]$ K. Karami, S. Ghaffari and M.M. Soltanzadeh, arXiv:1101.3240v1 [gr-qc] (2011).\\
$[38]$ M. R. Setare, {\it Phys. Lett. B} {\bf 641} 130 (2006).\\
$[39]$ G. Izquierdo and D. Pavon, {\it Phys. Lett. B} {\bf 633}
420 (2006).\\
$[40]$ A. Sheykhi, {\it J. Cosmol. Astropart. Phys.} {\bf 05} 019
(2009).\\
$[41]$ J. Zhou, B. Wang, Y. Gong and E. Abdalla, arXiv:0705.1264v2 [gr-qc] (2007).\\
$[42]$ H. M. Sadjadi, {\it Phys. Lett. B} {\bf 645} 108 (2007).\\
$[43]$ B. Wang, Y. Gong and E. Abdalla, {\it Phys. Rev. D} {\bf
74} 083520 (2006).\\

\end{document}